# Combined Iodine- and Sulfur-based Treatments for an Effective Passivation of GeSn Surface

L. Groell,[1] A. Attiaoui,[1] S. Assali,[1] and O. Moutanabbir[1,*]

[1] Department of Engineering Physics, École Polytechnique de Montréal, C. P. 6079, Succ. Centre-Ville, Montréal, Québec H3C 3A7, Canada

**TABLE OF CONTENTS:**

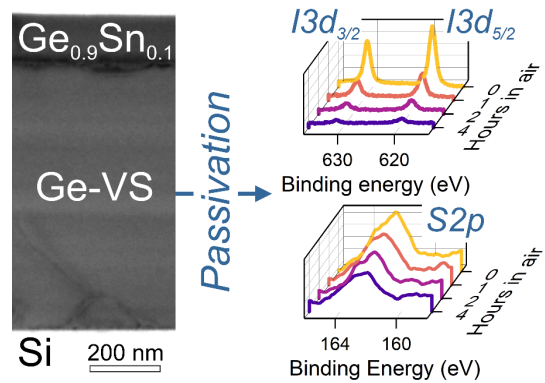

**ABSTRACT:**

GeSn alloys are metastable semiconductors that have been proposed as building blocks for silicon-integrated short-wave and mid-wave infrared photonic and sensing platforms. Exploiting these semiconductors requires, however, the control of their epitaxy and their surface chemistry to reduce non-radiative recombination that hinders the efficiency of optoelectronic devices. Herein, we demonstrate that a combined sulfur- and iodine-based treatments yields effective passivation of Ge and $Ge_{0.9}Sn_{0.1}$ surfaces. X-ray photoemission spectroscopy and *in situ* spectroscopic ellipsometry measurements were used to investigate the dynamics of surface stability and track the reoxidation mechanisms. Our analysis shows that the largest reduction in oxide after HI treatment, while $HF+(NH_4)_2S$ results in a lower re-oxidation rate. A combined $HI+(NH_4)_2S$ treatment preserves the lowest oxide ratio <10 % up to 1 hour of air exposure, while less than half of the initial oxide coverage is reached after 4 hours. These results highlight the potential of S- and I-based treatments in stabilizing the GeSn surface chemistry thus enabling a passivation method that is compatible with materials and device processing.

# INTRODUCTION

Developing all-group IV semiconductor light emitters and detectors is an attractive paradigm toward the monolithic integration of photonics and electronics. To this end, GeSn semiconductors have been in the limelight as their properties may pave the way for silicon-integrated infrared optoelectronics [1]. A tunable bandgap directness and energy in the extended short-wave and mid-wave infrared (e-SWIR and MWIR) wavelengths can be obtained by controlling the Sn content up to 30 at.% [2]. However, implementing this novel family of GeSn-on-silicon photonic platforms remains rather challenging from a materials standpoint as the equilibrium solubility of Sn in Ge is limited to ~1 at.%. Recent progress in controlling nonequilibrium growth kinetics allowed circumventing to a certain extent some of these thermodynamic constraints, thus limiting the phase separation and Sn segregation during the epitaxial growth of GeSn layers and heterostructures.[1-6] Consequently, the last few years witnessed the demonstration of a wide range of devices including photodetectors [7-13], optically- and electrically-pumped lasers [14-17], optical modulators [18], transistors [19-20], to name a few.

While significant efforts have been expended to control the growth and optoelectronic properties of GeSn [21-23], this material system still faces major challenges limiting the understanding and optimization of its fundamental properties. Controlling surface chemistry and passivation of dangling bonds are among these outstanding challenges that need to be addressed to improve the optoelectronic properties by reducing carrier traps and the associated nonradiative recombination channels. In low Sn content layers ($Ge_{0.97}Sn_{0.03}$), HF:HCl wet cleaning followed by $H_2O$ prepulsing showed a significant improvement in the current-voltage (C-V) characteristics as compared to untreated layers [24]. In the latter, a nearly flat high frequency response was observed, whereas HF:HCl-treated samples showed an enhanced inversion response with higher $C_{ox}/C_{min}$ ratio, steeper slope in depletion, and weak interface trap density response in strong inversion. This suggests that Cl-surface passivation largely improves GeSn electronic properties. Similarly, enhanced intrinsic conductance and hole mobility in Sulfur (S)-passivated $Ge_{0.83}Sn_{0.17}$ p-MOSFET were demonstrated by using HF and $(NH_4)_2S$ treatments before $HfO_2$ deposition [25]. Notwithstanding these early reports, little knowledge is currently available on the chemical stability of as-grown or treated GeSn surfaces [24-27]. While $(NH_4)_2S$ and $NH_4OH$ solutions alone do not remove Ge and Sn native oxides, combined treatments involving HF, HCl solutions strongly suppressed the oxides at the surface and provides a S-passivation layer [27]. However, the stability of the S-passivated GeSn surfaces is yet to be understood. Experimental studies on bulk Ge revealed that S-passivation obtained by aqueous ammonium sulfide solution, in contrast to chloride or hydride termination induced by HCl or HF treatments, can reduce durably the surface reactivity since the passivation overlayer shows an enhanced stability to oxidation in air for multiple days [28-30]. Other hydrohalic acid solutions such as HI and HBr were also shown to remove the entirety of Ge oxide and stabilize the surface beyond 12 hours of air exposure with a stable I- or Br-terminated surface [31-33]. Similar investigations on GeSn surfaces are still missing despite their importance to improve and optimize the optoelectronic properties of Sn-containing group-IV semiconductors.

With this perspective, herein we investigate the impact of wet treatments based on HI, HF, and $(NH_4)_2S$ chemicals on the stability of $Ge_{0.9}Sn_{0.1}$ and Ge surfaces. The dynamic change of the passivation layer is probed using X-ray photoemission spectroscopy (XPS) and spectroscopic ellipsometry, which provides a direct correlation with the increase in oxide thickness during surface reoxidation. The lowest amount of oxide at the surface is obtained after HI-based treatment, while the $(NH_4)_2S$-based treatment results in a lower re-oxidation rate. Moreover, we demonstrate that combining the S- and I- based treatments improves surface stability and slows down the oxide regeneration.

## EXPERIMENTAL SECTION

The investigated samples were grown on 4-inch B-doped (1-10 Ω·cm) Si (100) wafers in a low-pressure chemical vapor deposition (CVD) reactor using ultra-pure $H_2$ as a carrier gas, with monogermane ($GeH_4$), diluted at 10 % in $H_2$, and tin-tetrachloride ($SnCl_4$) as precursors, following a protocol similar to that described in Ref.[4]. The Ge virtual substrate (Ge-VS) on Si was grown using a two-temperature step process at 460 and 600 °C, followed by thermal cyclic annealing above 800 °C to improve the crystalline quality. The $Ge_{0.9}Sn_{0.1}$ growth on Ge-VS was conducted at 320 °C using a constant precursor supply with a $GeH_4/SnCl_4$ ratio of ~1700. In the cross-sectional transmission electron microscopy (TEM) image in Fig. 1a misfit dislocations are visible at $Ge_{0.9}Sn_{0.1}$/Ge-VS interface as a result of the lattice-mismatched growth. No threading dislocations are detected across the entire heterostructure at the TEM imaging scale.

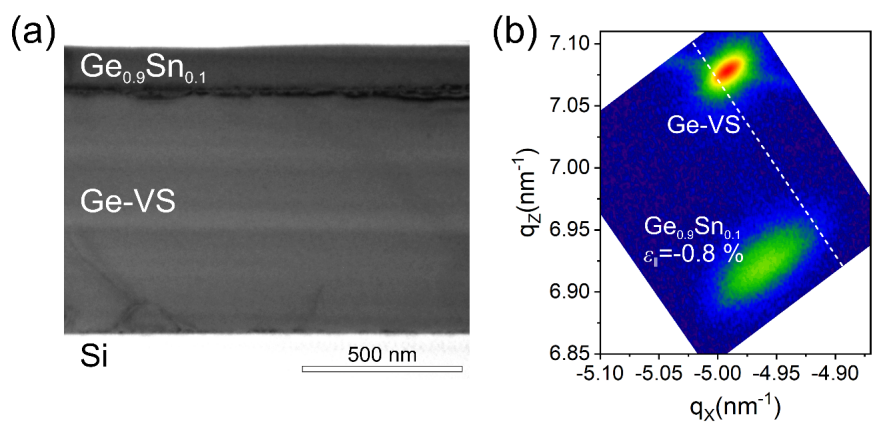

**Figure 1.** (a) Cross-sectional TEM image of the $Ge_{0.9}Sn_{0.1}$/Ge/Si multilayer heterostructure acquired along the [110] zone axis. (b) RSM recorded around the asymmetrical (224) reflection.

To decouple strain and composition in the material, reciprocal space mapping (RSM) around the asymmetrical (224) X-ray diffraction (XRD) peak are performed in Fig. 1b. Compressive in-plane strain $\varepsilon_{\parallel}$ =-0.8 % is estimated for the $Ge_{0.9}Sn_{0.1}$ layer, without showing phase separation and segregation of Sn at the surface [34]. XPS analysis was performed with a Mg Kα

X-ray source at 1253.6 eV operated under an anode bias of 15 kV with an emission current of 20 mA. Measurements were carried out at a take-off angle of 90° with respect to the sample surface plane at a base pressure of about 3.10$^{-9}$ Torr. Low-resolution survey scans were recorded at an energy step of 1 eV, a pass energy of 100 eV, and a dwell time of 100 ms. High-resolution spectra were collected with a resolution of 0.1 eV at a pass energy of 20 eV, and a dwell time of 100 ms. Peaks were charge referenced to the Ge3d$_{5/2}$ component at 29.4 eV [35]. Reference high-resolution XPS scans were collected on as-grown samples before any treatment. The quantification of XPS spectra was performed with relative sensitivity factors based on Wagner approximations [36]. The relative atomic composition is given in atomic %, which will be referred to here as %, with an accuracy of ±1 % (instrumental precision). The spectroscopic data were processed using the Avantage software package and the spectra were constrained and fit using a Smart background [36-37]. The handbook of X-ray Photoelectron Spectroscopy by Briggs et al. [35] has been used as the main reference for the binding energies along with NIST X-ray Photoelectron Spectroscopy Database [38]. To quantify the impact of surface treatments on GeO$_x$ and SnO$_x$ formation, the ratios of oxidized Ge and Sn atoms to the total of Ge or Sn atoms are calculated using the oxide ratios, $R_{GeOx}$ and $R_{SnOx}$:

$$R_{GeOx}(\%) = \frac{\sum A(GeOx)}{A(Ge) + \sum A(GeOx)} \qquad (1)$$

$$R_{SnOx}(\%) = \frac{\sum A(SnOx)}{A(Sn) + \sum A(SnOx)}, \qquad (2)$$

where A(GeO$_x$), A(SnO$_x$), A(Ge), A(Sn) are the normalized areas of Ge 3d$_{5/2}$ oxide(s), Sn 3d$_{5/2}$ oxide(s), Ge 3d$_{5/2}$, and Sn 3d$_{5/2}$ peaks, respectively. Only the 5/2 components of 3d peaks are considered and labelled on XPS figures for these ratios.

Room temperature variable angle spectroscopic ellipsometry (VASE or SE) measurements were conducted to assess the oxide overlayer thickness evolution as a function of time under ambient conditions. SE measurements were performed using a fast-dual rotating-compensator spectroscopic ellipsometer (RC2-XI, J. A. Woollam Co., Inc.). For the time-dependent measurements, the ellipsometric parameters ($\Psi$, $\Delta$) were measured before and 1 min after the treatment at one angle of incidence 75°, near the Brewster angle of Si, in steps of 1 nm throughout the spectral range from 300 nm to 2500 nm about every 10 s. For instance, when measurements are performed for 30 min, the number of obtained parameters becomes 2 × ($\Psi$, $\Delta$) × 2200 (the number of wavelengths) × 180 (the number of times) = ~792 000. Changes in Ge and GeSn surfaces, induced by different chemical treatment can be monitored in real time by measuring dynamically the changes in the imaginary part of the pseudo-dielectric function $\langle\varepsilon_2\rangle$ (see supplemental information S1), at the $E_2$ critical point (near 4.2 eV), which has been established to be very sensitive to the presence of surface overlayers and any possible surface irregularities such as surface roughness or adsorbents [39-41]. To model the oxide overlayer thickness, the Bruggeman effective medium approximation (BEMA) [42] was assumed, the oxide layer thickness and the void fraction are the model parameters for the epitaxial Ge, whereas three materials where considered for GeSn: the $GeO_2$ oxide, the $SnO_2$ oxide, and void. The BEMA is appropriate to microgeometries where the grains of various components are distributed, with no clear matrix component. Thus, the non-uniform coverage of the surface by $GeO_2$ and $SnO_2$ oxides [43] can be well represented with the aforementioned approximation. The optical properties of the thermally grown $GeO_2$ [44] and $SnO_2$ [45] layer were used from tabulated data in literature. In this work, given the scarcity of the optical properties of the native $GeO_2$ and $SnO_2$ oxides, their optical properties were thus approximated to that of a thermally grown uniform layer.

Three different wet chemical treatments (T1, T2, and T3) were investigated (see Table 1). T1 consists of a dip in HF (hydrofluoric acid) for 1 min followed by a dip in $(NH_4)_2S$ (ammonium sulfide) for 30 min. This treatment was selected as a reference since it was proposed as the most efficient for GeSn [27]. Since HI has been shown to be a good candidate for Ge passivation [31], we introduced a treatment T2 that is based on a 1 min HI (hydroiodic acid) dip to identify any similarities and differences between Ge and GeSn. Finally, T3 treatment is a combination of the two treatments T1 and T2, namely a dip in HI for 1 min succeeded by a dip in $(NH_4)_2S$ for 30 min. The rationale for this treatment is to investigate the combined efficiency of HI in removing the native oxide with the enhanced surface stability enabled by $(NH_4)_2S$. Each one of these treatment ends with a $N_2$ blow dry. The effects of each treatment were investigated for both epitaxial Ge and $Ge_{0.9}Sn_{0.1}$ samples.

|  | Treatment steps |
|---|---|
| T1 (S) | (1) HF 2% for 1 minute<br>(2) $(NH_4)_2S$ 20% for 30 minutes<br>(3) $N_2$ drying |
| T2 (I) | (1) HI 28.5% for 1 minute<br>(2) $N_2$ drying |
| T3 (I+S) | (1) HI 28.5% for 1 minute<br>(2) $(NH_4)_2S$ 20% for 30 minutes<br>(3) $N_2$ drying |

**Table 1.** Description of the wet treatments investigated in this study.

The time required to transfer a sample from the chemical solution into the XPS chamber was kept below 5 min, and it was below 2 min to start the SE *in situ* measurements, to avoid any substantial oxidation regrowth. A sequence of XPS analyses is carried out to follow the time-dependent evolution of the surface. Between each analysis the sample is unloaded and exposed to ambient air for a predetermined time to examine the oxide regrowth on $Ge_{0.9}Sn_{0.1}$ and Ge surfaces. The study flow is summarized in the chart displayed in Fig. 2.

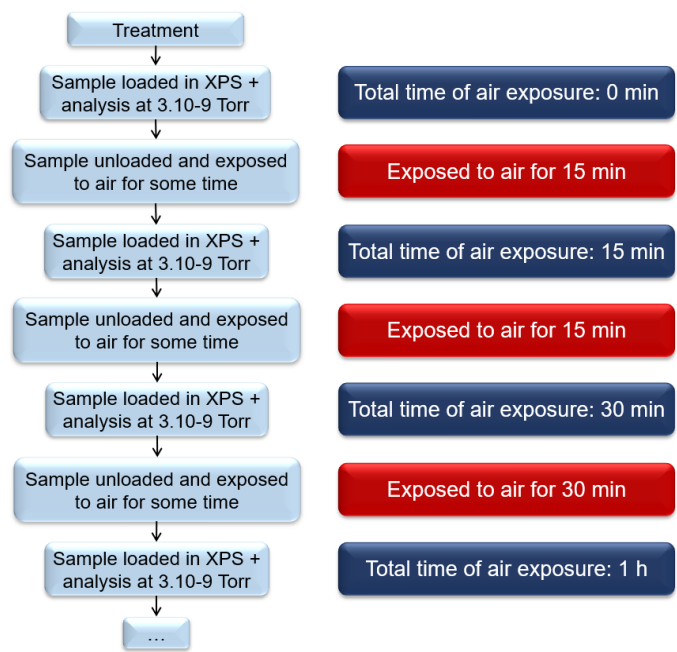

**Figure 2.** Process flow chart of the studies addressing the time-evolution of Ge and GeSn surfaces when exposed to ambient air after T1-T3.

## RESULTS AND DISCUSSION

**Oxide removal and surface passivation.** The XPS spectra for Ge3d (29.4 eV) and Sn3d (485 eV) peaks acquired for $Ge_{0.9}Sn_{0.1}$ before any treatment are shown in Fig. 3a-b. After T1 treatment, the intensities of the $GeO_x$ and $SnO_x$ peaks are strongly reduced in the Ge3d and Sn3d peaks detected (Fig. 3c-d). A reduction in the Ge oxide ratio at the surface from ~21 % (reference: 8.7 % of GeO + 12.5 % of $GeO_2$) to ~9 % is estimated from the fit of the XPS peaks. More precisely, two different suboxides with +1 (7.2 %) and +3 (1.9 %) oxidation states are present, with a chemical shift of +1.2 eV and +2.6 eV relative to $Ge3d_{5/2}$, respectively [29, 46].

Similar results were obtained for the epitaxially grown Ge layer on Si sample (see Supplemental Information S3). Interestingly, T1 treatment was found to have stronger effect on the Sn oxide peaks. We note a decrease from ~46 % (reference: 28.1 % of SnO + 18.3 % of SnO$_2$) to a mere 5.4 % with a +2 oxidation state[35, 47-48]. The enhanced removal of SnO$_x$ compared to GeO$_x$ is consistent with earlier observation using a (NH$_4$)$_2$S solution[25]. It is noteworthy that after T1 a S-related peak (S2p) is visible at 161.8 eV (Fig. 4a), which can be fitted using a single doublet model, thus indicating that only atomic S is bound to the surface[35].

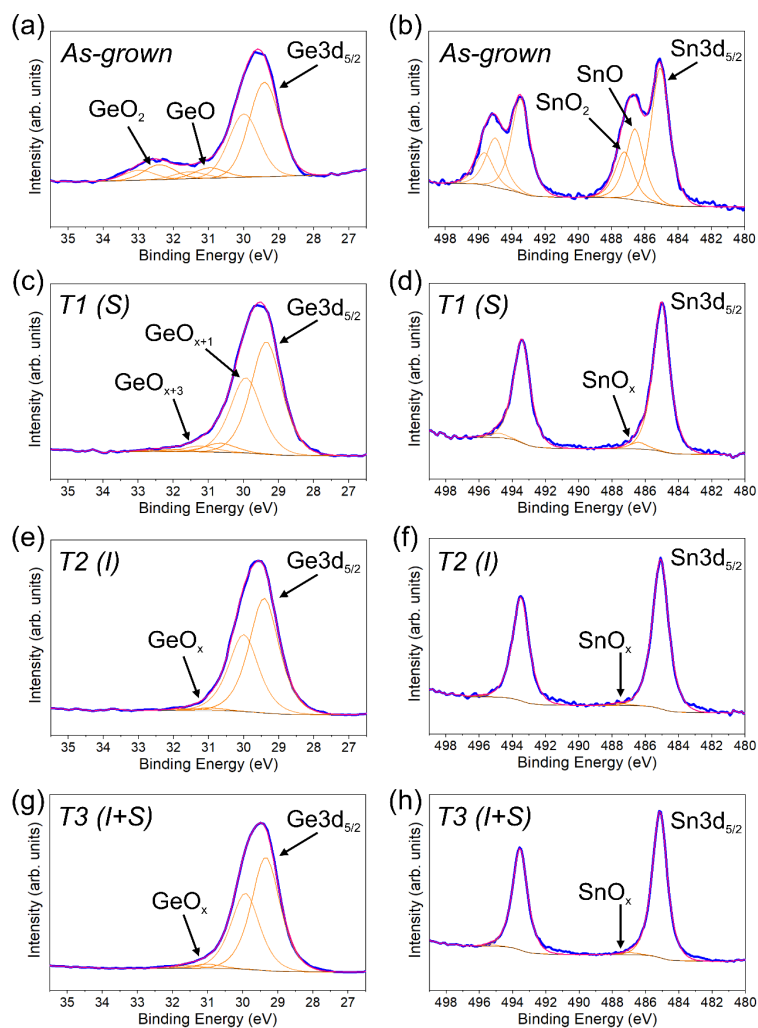

**Figure 3.** (a-h) Ge3d and Sn3d XPS peaks on the Ge$_{0.9}$Sn$_{0.1}$ surface in the as-grown sample (a-b) and after T1 (c-d), after T2 (e-f), and after T3 (g-h) treatments. Only the 5/2 components of the doublets are labelled.

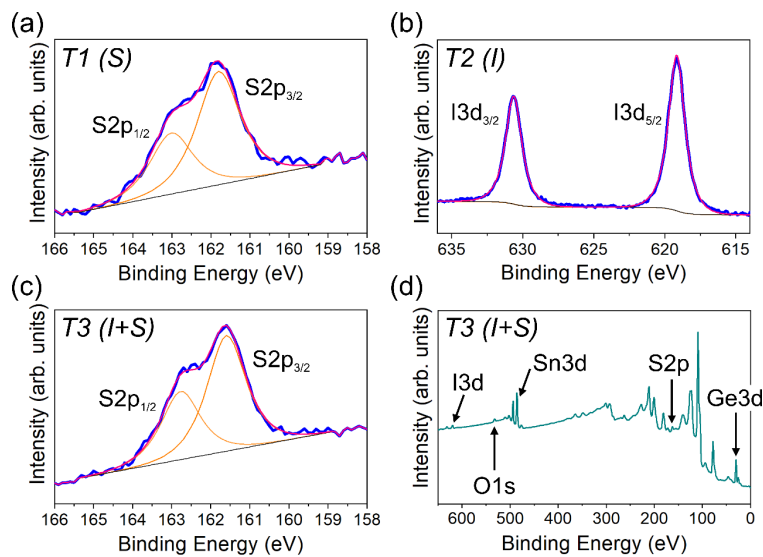

**Figure 4.** (a) S2p peak observed in Ge$_{0.9}$Sn$_{0.1}$ after T1. (b) I3d peak observed in Ge$_{0.9}$Sn$_{0.1}$ after T2. (c) S2p peak and (d) survey spectra showing a residual I3d peak for the Ge$_{0.9}$Sn$_{0.1}$ surface after T3.

We now turn our attention to the HI-based treatment (T2). The obtained results are displayed in Fig. 3e-f. The Ge oxide ratio is reduced from 21 % to 2.3 %, while the Sn oxides ratio decreases from ~46 % to 2.4 %. Only a negligible quantity of both Ge and Sn suboxides, with a +2 oxidation state [35, 38, 47-48], are left on the surface after T2. For comparison, HF-treated Ge surface is known to lose only partially Ge oxides [31-32, 49] and GeSn dipped in HF presents higher SnO$_x$ removal than GeO$_x$, while HI suppresses a larger amount and more equally both oxides. Therefore, we can conclude that HI treatment is highly effective in removing the native oxide on the GeSn surface. Moreover, a I3d (doublet) peak emerges at 619.2 eV (Fig. 4b), which indicates a surface passivation with one I species.

Next, we discuss the effects of the third treatment T3 exploring the combination of the enhanced removal of native oxide by HI-based T2 and the stable S-passivation layer of (NH$_4$)$_2$S-based T1 treatments. T3 consists of a HI dip for 1 min followed by a (NH$_4$)$_2$S dip for 30 minutes and N$_2$ drying, as outlined in Table 1. The acquired XPS spectra for T3-treated Ge$_{0.9}$Sn$_{0.1}$ are shown in Fig. 3g-h. Similar to T1 and T2, both GeO$_x$ and SnO$_x$ are strongly suppressed and the S-related peak (S2p) appears at 161.6 eV (Fig. 4c), while only a very low intensity I3d peak is visible (Fig. 4d). A very low surface oxidation is obtained, reaching 3.6 % for $GeO_x$ and 2.6 % for $SnO_x$, close to the observed behavior after T2.

**Time-dependent stability of GeSn treated surfaces.** To investigate the stability of the surface passivation layer in GeSn, the evolution of the $GeO_x$ and $SnO_x$ was monitored up to 4 hours after the three different treatments, as shown in Fig. 5 where error bars are related to the instrument's precision. Reference initial oxide ratios of $GeO_x$ ~21 % and $SnO_x$ ~46 % are considered for as-grown $Ge_{0.9}Sn_{0.1}$. Overall, regardless of the treatment the $GeO_x$ (Fig. 5a) tends to revert to its initial oxidation state (dashed line) faster than $SnO_x$ (Fig. 5b).

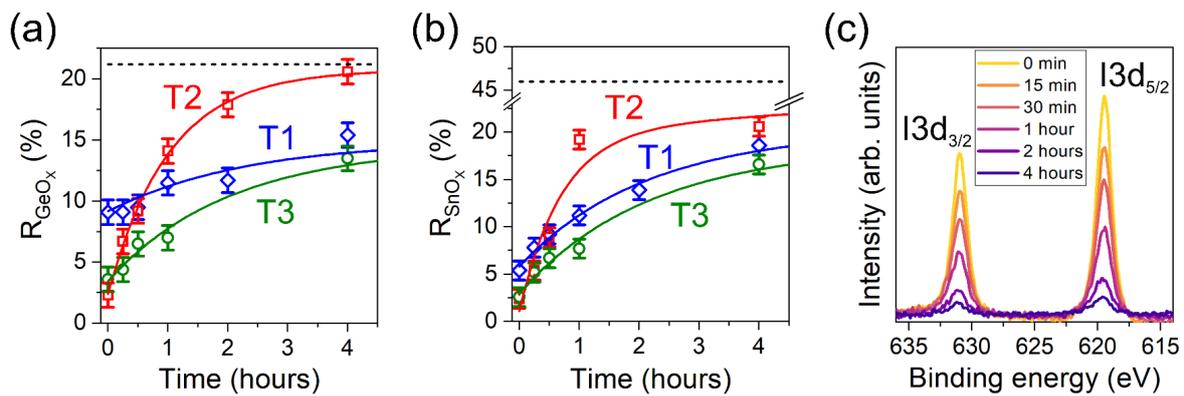

**Figure 5.** (a-b) $GeO_x$ (a) and $SnO_x$ (b) regeneration for the three treatments as a function of the exposure time to ambient air. A line is drawn for every treatment to follow the time-evolution. The horizontal dashed lines indication the initial oxide ratios before any surface treatment. (c) I3d peak recorded after variable exposure time of GeSn surface to ambient air.

T1 treatment shows the lowest oxide reduction of $GeO_x$ (~9 %) and $SnO_x$ (~5 %), however, this S-passivation provides the highest stability over time, with an increase from ~9 % to ~15 % for $GeO_x$ and from ~5 % to ~19 % for $SnO_x$ after 4 hours of exposure to ambient air. We also note that the S2p peak at 161.8 eV is still detected on GeSn surface even after more than 4 days of exposure to ambient air with a limited reduction in intensity. As discussed earlier in the text, T2 treatment presents the largest initial oxide reduction down to ~2 % for both $GeO_x$ and $SnO_x$, however, a faster oxide regrowth is observed with a steep increase in the $GeO_x$ from ~ 2 % to ~18 % in 2 hours (Fig. 5b). Similarly, the $SnO_x$ rapidly increases from the initial ~2 % to more than 15 % after only 1 hour of air exposure. Compared to T1 treatment, during the first hour, the oxide regrowth is at least 2 times faster after T2. We also note that the I3d peak intensity decreases over time (Fig.5c). After 4 hours of air exposure of the T2-treated GeSn surface, the I3d peak shows an 8-times smaller intensity than that measured immediately after the treatment, indicating that the I-passivation layer slowly degrades and oxide forms again. Furthermore, the Ge oxide ratio reverts to its initial value after 4 hours, whereas $SnO_x$ reaches ~60 % of its initial value. Interestingly, T3 outperforms both T1 and T2 treatments over a longer time scale, with a large oxide removal (similar to T2) that is combined with a slow reoxidation rate (similar to T1). After 4 hours of air exposure, the surface oxidation is only $GeO_x$~13 % and $SnO_x$~17 %, hence notably lower than the reference 21 % and 46 % in as-grown GeSn, respectively. Thus, T3 treatment allows for a high oxide reduction combined with a longer stability of the S-passivation surface.

The effectiveness of the three treatments was also investigated for epitaxial Ge surface and compared to GeSn, as shown in Fig. 6. For T1 treatment, the oxide regrowth on both Ge and GeSn surfaces is similar. In Ge, the $GeO_x$ slowly recovers to reach the initial 15 % coverage within 20h of air exposure (see Supplemental Information S4), while in GeSn the $GeO_x$ saturates at ~18 %, which slightly below its initial value (21 %). Previous reports showed an S-passivation of bulk Ge that is stable up to few days[28-30]. This discrepancy with our experimental observations could be attributed to a difference in surface energy of bulk and that of epitaxial Ge samples, as discussed later in the text. Note that even though T2 treatment is as efficient on Ge as on GeSn surface for native oxide removal, it provides a higher stability over time to Ge compared to GeSn surface (Fig. 6b). While earlier studies showed the potential of the HI treatment in passivating bulk Ge[31-32], the results shown here confirm the validity of this treatment to epitaxial Ge and GeSn. Finally, for T3 treatment we observe that the oxide regrowth follows a similar rate for both Ge and GeSn. However, after 4 hours the surface oxide coverage in Ge (11 %) almost recovers the as-grown value (14 %), while in GeSn (14 %) it remains well below its initial value (21 %). In addition, no residual I-related peak is found on the Ge surface and only the S2p peak (161.6 eV) is detected, while in GeSn the residual I peak (619.2 eV) is observed for 1 hour. We note that the order of dipping may play a role on which species is present on the surface, but this observation could imply that iodine is more likely to remain bonded to Sn atoms on the GeSn surface.

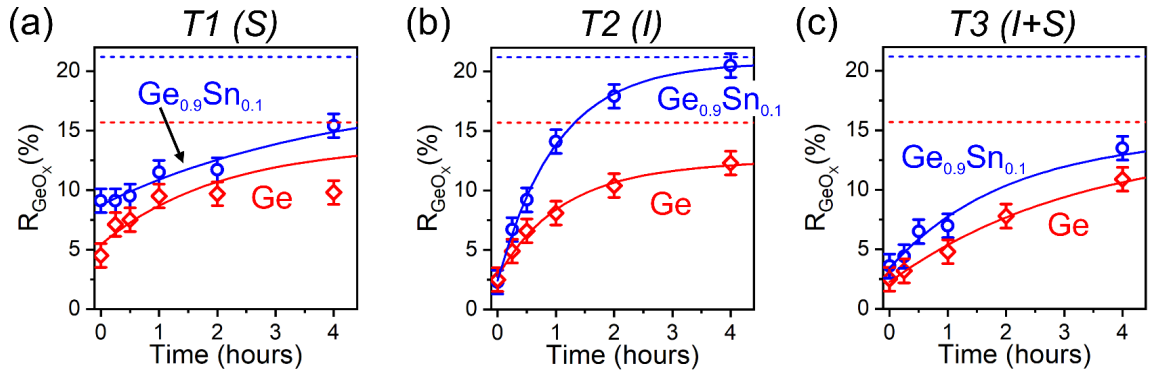

**Figure 6.** (a-c) GeO$_x$ ratio evolution for epitaxial Ge (in red) and GeSn (in blue) surfaces after T1 (a), T2 (b), and T3 (c). The horizontal dashed lines represent the initial Ge oxide ratio for epitaxial Ge (in red) and GeSn (in blue). The solid lines are guide for the eye.

**Unravelling surface instability by *in situ* spectroscopic ellipsometry.** To further investigate the stability over time of the passivation treatments spectroscopic ellipsometry (SE) measurements were performed. The imaginary part of the dielectric function $\langle \varepsilon_2 \rangle$ for the Ge and GeSn samples are displayed in Fig. S1 of the Supplemental Information. Larger values of $\langle \varepsilon_2 \rangle$ generally correlates with a lower surface roughness and presence of particles on the surfaces [39-41]. Overall, a decrease in $\langle \varepsilon_2 \rangle$ with air exposure time is observed in all samples and indicates that an oxide overlayer is growing on the surface. The highest $\langle \varepsilon_2 \rangle$ for both epitaxial

Ge and GeSn samples is obtained for T3, which agrees well with the XPS data showing the lowest oxidation over 4 hours of air exposure. Next, we define the relative changes of the thickness, $\Delta d/d$ as a function of the exposure time to air as

$$\frac{\Delta d}{d} = \frac{d_{Ti} - d_{Ti}^0}{d_{Ti}^0}, \qquad (3)$$

where $d_{Ti}$ is the surface overlayer thickness after the Ti treatment (i=1,2,3) extracted from the BEMA SE model, and $d_{Ti}^0$ is the surface overlayer thickness before any treatment. The thicknesses of the latter are presented in Table 2. Fig. 7 shows the relative change of the effective overlayer thickness on Ge and on GeSn after T1, T2 and T3 treatments. The zero in the vertical axis refers to the untreated surface (as-grown sample), while negative values correspond to a decrease in the thickness of the oxide layer. After wet treatment, the lowest $\frac{\Delta d}{d}$ values are obtained, followed by a monotonous increase indicating that surface oxidation occurs. The profile (black dashed lines) for each treatment was fitted to a logistic growth model to guide the eye. The thickness error bars are related to the error associated with the SE model. The grey vertical line in Fig. 7 corresponds to the same time of air exposure as in the XPS measurements. T3 leads to a relative thickness change that is smaller than in T1 and T2 treatments on both Ge (Fig. 7a) and GeSn (Fig. 7b) even after few hours of air exposure. This translates to a smaller surface overlayer thickness in T3 with regards to T1 and T2, thus consolidating the efficacy of the T3 treatment. A qualitative agreement between the SE and XPS measurements is obtained for the oxide reduction and re-growth rate especially for sulfur-based treatments, for both GeSn and Ge semiconductors. However, a larger than expected relative change is measured in the first 30 min after the treatment T2 for GeSn sample. While the XPS data indicate that T2 treatment is the most effective in removing the oxide during the first 30 min of exposure to ambient air, SE studies reveal that the same treatment also yields the largest oxide thickness within the first minutes of air exposure. The most likely explanation for this discrepancy between XPS and SE results for T2 could arise from the fact that the structural and optical properties of the Iodine passivation layer are largely unknown at this stage, which would affect our precision in estimating the thickness I-layer thickness.

|         | $d_{Ge}$ (nm) | $d_{Ge0.9Sn0.1}$ (nm) |
|---------|---------------|------------------------|
| T1 (S)  | 1.46±0.10     | 1.65±0.20              |
| T2 (I)  | 1.58±0.11     | 1.85±0.25              |
| T3 (I+S)| 1.28±0.10     | 1.58±0.15              |

**Table 2.** Overlayer-thickness $d$ (i.e., native oxide) measured before any treatment for the Ge and $Ge_{0.9}Sn_{0.1}$ layers.

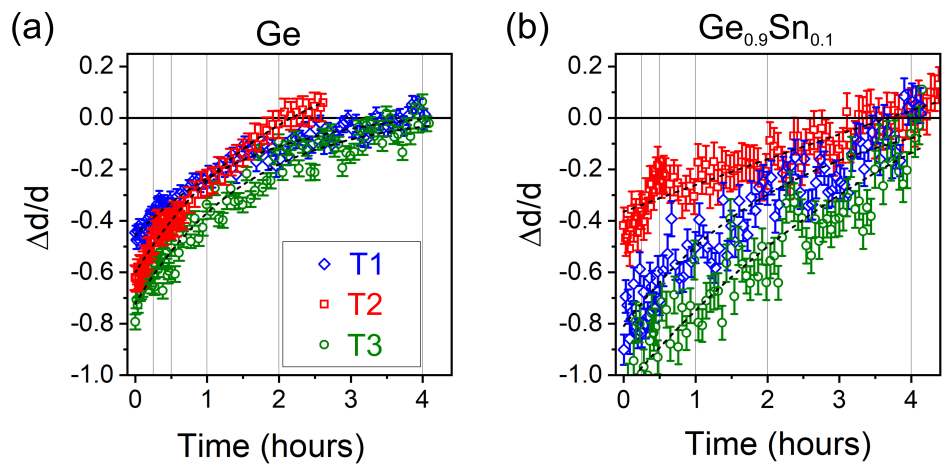

**Figure 7.** (a-b) Effective thickness estimation from SE for Ge (a) and $Ge_{0.9}Sn_{0.1}$ (b) for all treatments (T1, T2 and T3). The zero in the vertical axis corresponds to the as-grown overlayer thickness (native oxide). The black dashed lines are a sigmoid growth fit to guide the eye. The vertical grey lines correspond to the XPS time of measurement shown in Fig. 4.

**Stability of the passivated surface.** The data presented above for GeSn indicate that after T1, the $GeO_x$ and $SnO_x$ regrowth is the slowest and a stable S passivation is maintained for at least a few hours. The detected S peak is attributed to a slowdown of the GeSn surface oxidation [27], thereby creating GeSn-S bonds which passivate the highly reactive dangling bonds and stabilizes the surface. In the absence of a passivation layer, the oxidation of GeSn is extremely fast upon exposure to air, as recently discussed by Raynal et *al*. for $Ge_{0.85}Sn_{0.15}$, where in less than 15 min $R_{SnOx}$ ~90 % and $R_{GeOx}$ ~13 % were measured [26]. On the contrary, our results show much lower ratios of $R_{GeOx}$ ~9.1 % and $R_{SnOx}$ ~5.4 % 15 min after the S-based wet treatment T1, which hints to the effective S-based passivation of GeSn surface. Moreover, we noted that the I-passivation desorbs faster than the S-passivation, thus suggesting weaker I-bonds compared to S-bonds. The lower bond dissociation energy for I (E(Ge-I) = 268 kJ/mol, E(Sn-I) = 235 kJ/mol) compared to S (E(Ge-S) = 534 kj/mol, E(Sn-S) = 467 kJ/mol) could explain the short-lasting passivation of T2 compared to T1 [50]. We also highlight that GeSn tends to form thermodynamically more stable bonds with oxygen (E(Ge-O) = 658 kj/mol, E(Sn-O) = 528 kj/mol) [50]. Consequently, when $H_2O$ and $O_2$ molecules get closer to the GeSn surface new O bonds are likely to form. However, other mechanisms could be involved as solid Iodine is known to have a high volatility and can sublimate when pressure decreases under 0.027 kPa [51]. Thus, newly formed I-layer could be subjected to desorption.

Our results demonstrate that T3 treatment combines the key advantages of T1 and T2 treatments, allowing a large oxide removal with HI along with improved stability of the surface due to the S-passivation. More precisely, T3 outperforms T1 and T2 on both Ge and GeSn according to XPS and SE data. Based on the experimental results outlined above, we propose the following interpretation to elucidate the behavior of the GeSn surface chemistry following the three treatments. After the HI dip, the surface of GeSn becomes oxide-free and most of the dangling bonds are saturated with I atoms. The newly formed I bonds are expected to prevent Ge and GeSn surfaces from re-oxidation more effectively than H bonds that form after a HF dip, since H bonds are unstable on the Ge surface [31, 52-55]. Subsequently, when the sample is immersed in the (NH$_4$)$_2$S solution, due to the higher electronegativity of S with respect to I, I bonds are replaced by S bonds on the surface that are thermodynamically relatively more stable. The bond dissociation energies mentioned above [50] supports this mechanism. Besides, we observed that after any of the treatment, GeSn treated surface is more oxidized than treated epitaxial Ge surface. One may invoke the following two points as a plausible explanation. First, since Sn is more electropositive than Ge, the reactivity of the GeSn surface upon oxygen exposure is expected to be higher than in Ge [26]. Second, the lattice-mismatched growth of GeSn/Ge-VS on Si results in a <110>-oriented cross-hatched pattern at the surface [56], with hills and valleys that could affect the stability and reactivity of the surface as a result of local changes in surface energy [57-59], potentially triggering faster oxidation. It is also important to note that the difference in stability of the I- and S-passivation layers on Ge and GeSn surfaces after different treatments compared with previous studies could arise from the difference in surface roughness between the two sets of samples. For instance, we noticed that there are major differences in surface stability between bulk and epitaxial Ge after the same treatment T2. In fact, Kim *et al.* reported no reoxidation of the surface after 12 hours [31], whereas in our study on epitaxial Ge a ~12 % of GeO$_x$ is estimated after 12 hours (see Supplemental Information S4). It is worth noting that previous Ge passivation studies were conducted on bulk Ge wafers [30-33, 49, 52-53, 60], which have a surface roughness that is one order of magnitude lower compared to epitaxial Ge on Si. Furthermore, threading dislocations (density typically above $5\times10^6$ cm$^{-2}$) can commonly reach the surface in epitaxial Ge on Si [61], leading to densities that are orders of magnitude higher than in bulk Ge. The larger number of defects combined with the higher roughness could increase the reactivity of the Ge and GeSn surfaces, thus reducing post-treatment surface stability. The results shown here clearly indicate that a stable passivation of Ge and GeSn can be achieved for a time scale of a few hours, which is relevant for device fabrication. One would expect higher surface stability, in controlled environment where ambient conditions can be monitored (humidity, temperature, etc.).

## CONCLUSIONS

In summary, we investigated the behavior of GeSn and Ge surface chemistry after native oxide removal using different wet chemical treatments. The lowest amount of oxide at the surface is obtained after HI-based treatment, while the $(NH_4)_2S$-based treatment results in a lower re-oxidation rate. A combined HI+$(NH_4)_2S$ treatment results in the highest oxide reduction and longest stability, with a surface oxide content remaining lower than 10 % after 1 hour of exposure to ambient air. The correlation with spectroscopic ellipsometry measurements confirms the monotonous increase in the surface oxide after the wet conditioning process. Moreover, the oxide regrowth is further suppressed even at longer times of exposure to air, reaching less than half of the initial oxide coverage after 4 hours. These results show that the combined S- and I-based treatment is effective passivation method to stabilize GeSn surface, thus providing a reliable path to improve GeSn-based photonic and opto-electronic devices. Metallic Ohmic contacts with a reduced resistance could be achieved by passivating dangling bonds at the GeSn surface as a result of Fermi level depinning [62-63], which is critical to develop photodetectors or photoconductors. Moreover, the reduced density of interface traps on a passivated surface would improve the electrical performances of GeSn-based transistors [27, 64].

## AUTHOR INFORMATION

Corresponding Author:

*E-mail: oussama.moutanabbir@polymtl.ca

Notes

The authors declare no competing financial interest.

# ACKNOWLEDGEMENTS

The authors thanks J. Bouchard for the technical support with the CVD system, B. Baloukas for support with the spectroscopic ellipsometry measurement. O.M. acknowledges support from NSERC Canada (Discovery, SPG, and CRD Grants), Canada Research Chairs, Canada Foundation for Innovation, Mitacs, PRIMA Québec, and Defence Canada (Innovation for Defence Excellence and Security, IDEaS).

# SUPPLEMENTAL INFORMATION

Supplemental information includes the time evolution of the imaginary part of pseudo-dielectric function $\langle\varepsilon_2\rangle$ for Ge and GeSn, the dynamic AFM maps for Ge before and after treatment, XPS results of Ge surface treated by T1, T2 and T3, and evolution of Ge and GeSn oxidation beyond 4h of exposure to air after T1 and T2.

# Supplemental Material:

# Combined Iodine- and Sulfur-based Treatments for an Effective Passivation of GeSn Surface

L. Groell,[1] A. Attiaoui,[1] S. Assali,[1] and O. Moutanabbir[1,*]

[1] Department of Engineering Physics, École Polytechnique de Montréal, C. P. 6079, Succ. Centre-Ville, Montréal, Québec H3C 3A7, Canada

## Contents

## S1. *In-situ* spectroscopic ellipsometry

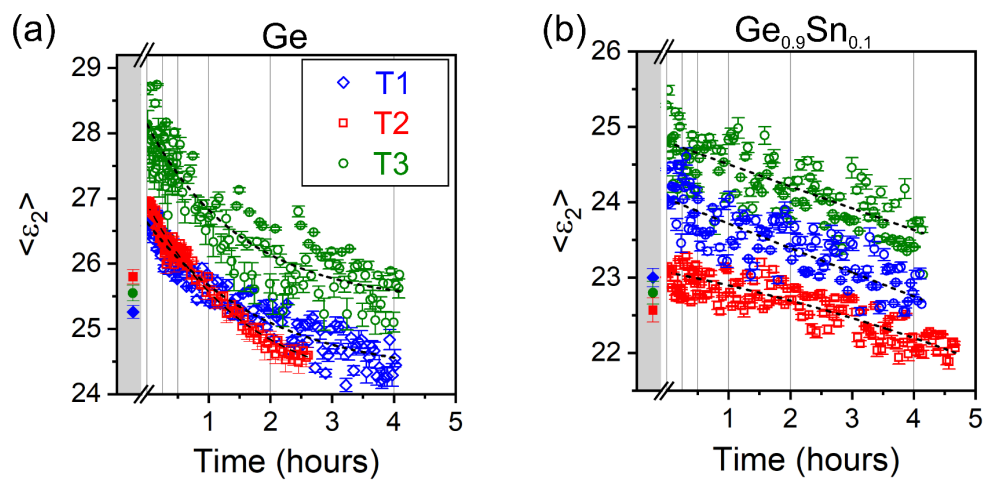

**Figure S1.** Real time $\langle \varepsilon_2 \rangle$ data for the T1, T2 and T3 treated Ge (left) and GeSn (right) surfaces, at ambient, near the $E_2$ critical-point energy (around 4.2 eV). The black dashed lines are a sigmoid growth fit to guide the eye. An angle of incidence (AOI) of 75° at 4.2 eV was used.

The imaginary part of the pseudo-dielectric function $\langle \varepsilon_2 \rangle$ is very sensitive to the surface changes at the $E_2$ critical point (4.2 eV), that is because the probe depth is of about 10 nm. For example, a reduction of the thickness of the surface overlayer or surface roughness (RMS) leads to increasing $\langle \varepsilon_2 \rangle$. The differences observed in the pseudo-dielectric function in Ge (Figure S1) are attributed to void fraction variations, and not RMS fluctuations (S2), between the Ge film and/or the presence of a surface layer, assumed as non-crystalline $GeO_2$ oxide. A comparison of the pseudo-dielectric function $\langle \varepsilon_2 \rangle$ spectra at 4.2 eV, shows an average (between Ge and GeSn values) increase of ~5 %, ~3 %, and ~9 % of $\langle \varepsilon_2 \rangle$ value at $E_2$ after the treatment T1, T2 and T3 respectively, for Ge and GeSn. This figure highlights the changes in $\langle \varepsilon_2 \rangle$ in real time under ambient conditions for Ge and GeSn for the three different chemical treatments.

## S2. AFM maps for Ge before and after treatment

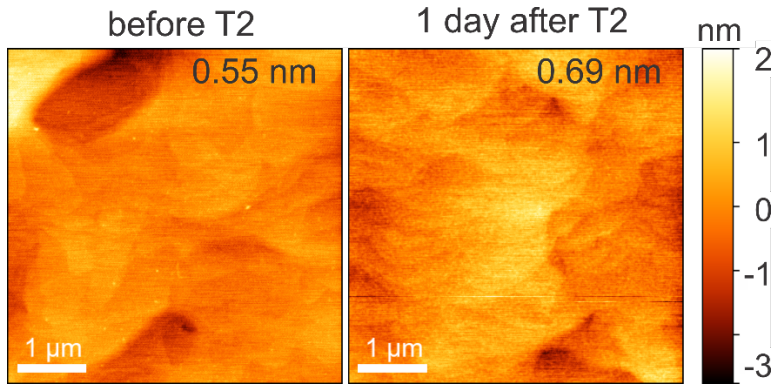

**Figure S2.** 5 µm × 5 µm AFM map of the Ge film before the T2 treatment and 1 day after the T2 treatment. The RMS is indicated in the maps and the change is very small.

The 5 µm × 5 µm atomic force microscopy (AFM) images (S2) measured before and 1 day after the T2 treatment for the Ge film indicates that the RMS fluctuation are less than 1.5 Å which consolidate the hypothesis of void fraction variation for the differences in $\langle \varepsilon_2 \rangle$.

## S3. Oxide removal on epitaxial Ge

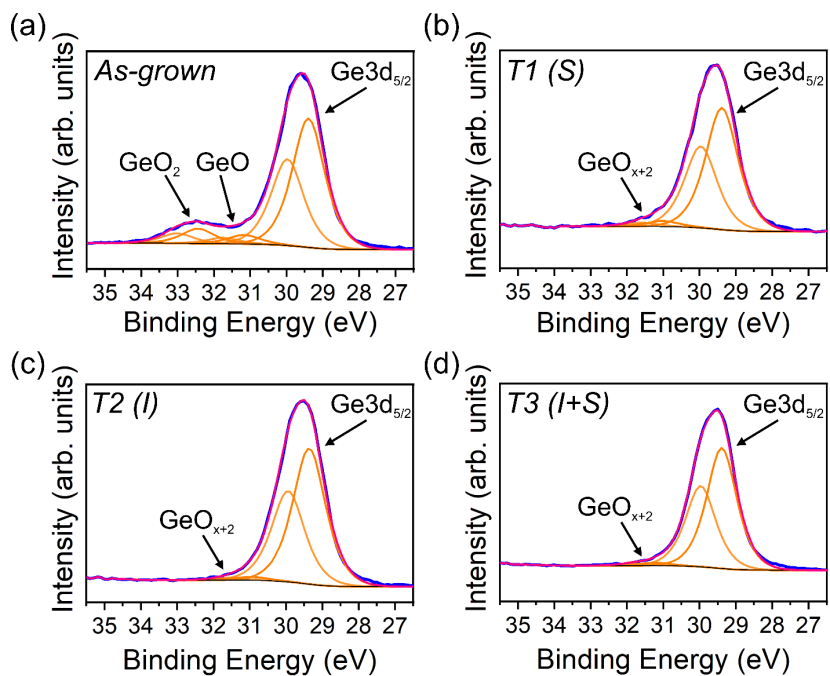

**Figure S3.** Ge3d on epitaxial Ge surface (a) reference before any treatment, (b) after T1, (c) after T2, (d) after T3. Only the 5/2 components of the doublets are labelled.

In a similar way that GeSn, after T1-3, intensities of GeOx are strongly reduced on Ge surface. The initial oxide ratio on epitaxial Ge surface was of about ~16 % (around 10% and 6% attributed to +4 and +2 oxidation states respectively). After T1, it is reduced to ~ 4 % and only some suboxides are presents, they are attributed to +2 oxidation state since they present a chemical shift of 1.6 eV with respect to $Ge3d_{5/2}$. Oxide removal is even larger for T2 and T3, oxide ratio is reduced to only 2.3% the treatment with a +2 oxidation state as well.

## S4. Evolution of Ge and GeSn oxidation beyond 4 h of air exposure

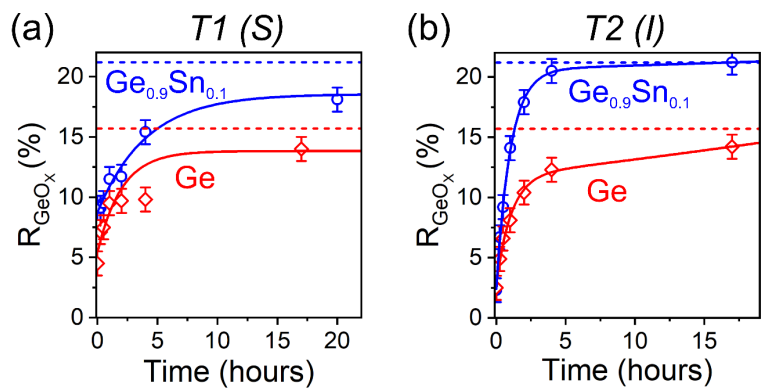

**Figure S4.** Oxide regrowth on Ge and GeSn surfaces after (a) T1 and (b) T2.